\documentclass[twocolumn,aps,prl,groupedaddress]{revtex4}
\usepackage{graphicx,wrapfig,amsmath,amssymb,enumitem,upgreek,bbm,mathtools,amsfonts,physics,multirow,color,caption,subcaption,hyperref,xcolor,lineno}
\usepackage[normalem]{ulem}

\renewcommand{\vec}[1]{\oldbm{#1}}

\renewcommand{\vec}[1]{\boldsymbol{#1}}

\def\bb{{\vec b}}
\def\bk{{\vec K}}
\def\bk{{\vec k}}

\def\bA{{\vec A}}

\def\bq{{\vec q}}

\def\bv{{\vec v}}

\def\bG{{\vec G}}

\def\bn{{\vec n}}
\def\bm{{\vec m}}

\def\br{{\vec r}}

\def\tr{\mathop{\mathrm{tr}}}

\def\H{\mathcal{H}}

\def\diag{{\rm diag}}

\def\tr{{\rm tr}}

\def\SU{{\rm SU}}

\newcommand{\clr}{\color{black}}

\newcommand{\beq}{\begin{equation}}
\newcommand{\eeq}{\end{equation}}
\newcommand{\beqarray}{\begin{eqnarray}}
\newcommand{\eeqarray}{\end{eqnarray}}

\begin{document}

\title{
From electrons to baby skyrmions in Chern ferromagnets: A topological mechanism for  spin-polaron formation in twisted bilayer graphene
}

\author{Eslam Khalaf and Ashvin Vishwanath}
\affiliation{Department of Physics, Harvard University, Cambridge, MA 02138}

\begin{abstract}
 The advent of Moir\'e materials has galvanized interest in the nature of charge carriers in topological bands. In contrast to conventional materials where charge carriers are electron-like quasiparticles,  topological bands allow for more exotic possibilities where charge is carried by nontrivial topological textures, such as skyrmions. 
However, the real space description of skyrmions is ill-suited to address the limit of small or `baby' skyrmions which consist of an electron and a few spin flips. Here, we study the formation of the smallest skyrmions -- spin polarons, formed as bound states of an electron and a spin flip -- in Chern ferromagnets. We show that, quite generally, there is an attraction  between an electron and a spin flip that is purely topological in origin and of $p$-wave symmetry,  which promotes the formation of spin polarons. Applying our results to the topological bands of twisted bilayer graphene, we identify a range of parameters where  spin polarons are formed and are lower in energy than electrons. In particular, spin polarons are found to be energetically cheaper on doping correlated insulators at integer fillings towards charge neutrality, consistent with the absence of quantum oscillations and the rapid onset of flavor polarization (cascade) transition in this regime. Our study sets the stage for pairing of spin polarons, helping  bridge skyrmion pairing scenarios  and momentum space approaches to  superconductivity.

\end{abstract}

\maketitle

\emph{\bf Introduction}--- There has been much recent interest in narrow Chern bands that  spontaneously develop ferromagnetic order, following their appearance in a variety of Moir\'e materials \cite{PabloSC,PabloMott, Dean-Young, Efetov, YoungScreening, EfetovScreening,  CascadeShahal, CascadeYazdani, RozenEntropic2021, CaltechSTM, RutgersSTM, ColombiaSTM, PrincetonSTM, SharpeQAH, YoungQAH,YankowitzMonoBi1, YankowitzMonoBi2, YoungMonoBi, TDBG_Pablo,TDBGexp2019,IOP_TDBG,YankowitzTDBG,KimTDBG,he2021chiralitydependent,Lee2019,Zhang2018, ABCMoire, BalentsReview}. The bands of magic-angle twisted bilayer graphene (MATBG) can, for example, be viewed as complementary Chern bands residing on opposite sublattices \cite{Zou2018,Tarnopolsky,Zhang2018,XieMacdonald, KIVCpaper}, where spin, valley and sublattice polarization leads to Chern insulators (and even fractional Chern insulators \cite{XieFCI}) as observed in experiment \cite{YoungQAH, EfetovQAH}. A key question is, what is the nature of charge carriers associated with doping these generalized Chern ferromagnets?  The answer will have implications for the entire phase diagram of various Moir\'e materials, and could hold the key to explaining mysteries such as the doping dependence and origin of superconductivity in MATBG and related structures. The simplest example of a Chern band, Landau levels, have  previously been shown to exhibit a ferromagnetic  ground state at unit filling \cite{Sondhi, MoonMori}. Despite the simplicity of the ground state, charge excitations can be very non-trivial. In addition to single electron quasiparticles corresponding to adding an electron with a reversed spin, this system also hosts charged skyrmions -  smooth  texture of the ferromagnetic order that carry an electric charge proportional to their topological winding \cite{Sondhi,KaneLee}. In the presence of spin rotation symmetry, it was shown that  big skyrmions are in fact the cheapest excitation\cite{Sondhi,MoonMori}, and they expand to reduce Coulomb repulsion, meaning that they incorporate many spin flips. On the other hand, with spin anisotropy or Zeeman field, the skyrmions shrink to `baby skyrmions' \cite{BabySkyrmions}, corresponding to an electron dressed by a few spin flips \cite{PalaciosFertig}. The smallest nontrivial limit of a skyrmion is a quasiparticle bound to a single spin flip, i.e. a {\em spin-polaron}. (For a discussion of spin-polarons in \emph{non-topological} ferromagnets, see e.g. Ref.~\cite{NagaevPolarons, PolaronShastry, PolaronReview, PolaronBerciu}).

In this paper, we will study the nature of charge excitations in Chern ferromagnets, and apply the results to models of twisted bilayer graphene. Although the spectrum of single-particle excitations in such systems has been obtained from self-consistent Hartree-Fock studies \cite{XieMacdonald, GuineaHF, ShangNematic, KIVCpaper} whose results are exact in certain limits \cite{KangVafekPRL, TBGV}, the existence of other low-lying {charged} excitations, e.g. skyrmions or spin-polarons, implies that a single-quasiparticle based description is insufficient to capture the physics on doping correlated insulators. Such unconventional excitations were proposed by the authors and co-workers to play a crucial role in the `skyrmion mechanism' of superconductivity \cite{SkPaper,chatterjee2020skyrmion}.

Although skyrmions can be smoothly shrunk to electrons, the standard description of the two excitations could not be more different: the former is a real space texture whose energy is computed using effective field theory \cite{MoonMori, Sondhi} or real space variational methods \cite{SkyrmionsWithoutSigmaModel, Fertig1994, Fertig1997} while  the latter is a momentum eigenstate whose energy is obtained from momentum-space Hartree-Fock. A small skyrmion, and in particular a spin polaron, is more naturally described as an electron dressed by spin flips \cite{PalaciosFertig} rather than as a real space texture. In fact, variational estimates for small skyrmions based on the latter vastly overestimates their energy and obscures the relation to single particle excitations. On the other hand, thinking of small skyrmions as electrons dressed by spin flips poses a puzzle: how is information about band topology, necessary for charged skyrmions, incorporated in the dressed electron picture, i.e. how does a localized excitation in momentum space detect the topology of the entire band?

In this work, we take the first step towards a momentum space characterization of non-trivial charge excitations in Chern ferromagnets by studying the formation of the smallest skyrmions, the spin polaron, consisting of an electron dressed by one spin flip. We show that the matrix elements of an arbitrary density-density interaction $V_\bq$ between an electron-magnon state with magnon momentum $\bq$ and one with momentum $\bq'$ is $\propto i \bq \wedge \bq' \frac{2\pi C}{A_{\rm BZ}} V_{\bq - \bq'}$ at small $\bq$ and $\bq'$, where $C$ is the Chern number. Remarkably, this implies there is an {\em attractive} interaction between an electron and a spin-flip for any repulsive interaction $V$,  which takes place in the ($p_x + i p_y$)-wave channel, as a consequence of the band topology.

We further investigate the conditions under which such attractive interaction leads to the formation of a bound state, i.e. a spin polaron. We consider a class of continuum models that interpolates between the lowest Landau level (LLL) and the Chern bands of twisted bilayer graphene (TBG) enabling us to tune band topology, geometry and dispersion quasi-independently. 
To solve this problem, we exploit the fact that the Hilbert space of an electron + a single spin flip only scales as $N^2$ for a system with $N$ unit cells which allows us to solve this problem for relatively large system sizes (up to $13 \times 13$) using exact diagonalization. Our results can be summarized as follows: (i) In the limit of vanishing quasiparticle dispersion (i.e. the non-interacting dispersion + the interaction generated dispersion), we find that the spin polaron is always lower in energy than the electron, with its energy increasing as the Berry curvature gets more concentrated. (ii) The existence of the spin polaron as a bound state is very sensitive to band topology and it is lost if we drive a phase transition to trivial Chern bands. 
(iii) We find that there is a critical value for the effective mass of the quasiparticle bands, beyond which the single electron is lower in energy than an electron dressed with a spin flip. In this limit, although the spin polaron does not exist as a stable bound state, it can still influence the physics as a resonance.
Our results serve as a bridge between momentum space single particle excitations and real space skyrmion excitations with the energy of the spin polarons computed here providing a strict upper bound on the energy of skyrmion excitations. Furthermore, our results show that a description in terms of single-particle excitations, \emph{even when they are `exact'}, is generally incomplete to understand the physics of charge doping in a Chern ferromagnet. 
At the end, we discuss the implications of these results for the phenomenology of TBG.

\emph{\bf General formalism}--- We consider the Hamiltonian of a density-density interaction $V_\bq$ projected onto a pair of $\SU(2)$-symmetric bands with single particle dispersion $\epsilon_0(\bk)$ and wavefunctions  $|u_\bk \rangle$:
\begin{equation}
    \H = \sum_{\bk,\sigma = \uparrow,\downarrow} c_{\bk,\sigma}^\dagger \epsilon_0(\bk) c_{\bk,\sigma} + \frac{1}{2A} \sum_\bq V_\bq \delta \rho_\bq \delta_{-\bq},
    \label{Ham}
\end{equation}
where $\delta \rho_\bq = \rho_\bq - \bar \rho_\bq$ is the projected density measured relative to a certain reference chosen such that the interacting piece of the Hamiltonian annihilates the ferromagnetic state at half-filling (see Ref.~\cite{KIVCpaper} for details). The projected density operator is given by
    \begin{equation}
        \rho_\bq = \sum_\bk c^\dagger_{\sigma, \bk} c_{\sigma, \bk + \bq} \lambda_\bq(\bk), \qquad \lambda_\bq(\bk) = \langle u_\bk| u_{\bk + \bq} \rangle
    \end{equation}
    
If the bare dispersion $\epsilon_0$ is sufficiently small, the ground state of the Hamiltonian (\ref{Ham}) is a ferromagnet, with total spin $S = \frac{N}{2}$, which we can choose to have $S_z = -\frac{N}{2}$ leading to 
\begin{equation}
    |\downarrow \rangle = \prod_\bk  c_{\downarrow, \bk}^\dagger |0 \rangle, \qquad \delta \rho_\bq |\downarrow \rangle = 0, \, \forall \bq
\end{equation}

Single particle excitations with charge $\mp e$ are given by $|\bk \rangle_e = c_{\uparrow,\bk}^\dagger |\downarrow \rangle$ and $|\bk \rangle_h = c_{\downarrow,\bk} |\downarrow \rangle$, respectively. The state $|\bk \rangle_{e/h}$ has $S_z = -\frac{N-1}{2}$ and total spin $S = \frac{N-1}{2}$. Its energy can be computed exactly using the commutation relations
\begin{equation}
    [\delta \rho_\bq, c_{\sigma, \bk}^\dagger] = \lambda_{-\bq}^*(\bk) c^{\dagger}_{\sigma, \bk - \bq}, \quad [\delta \rho_\bq, c_{\sigma, \bk}] = - \lambda_{\bq}(\bk) c_{\sigma, \bk + \bq}
    \label{Commutators}
\end{equation}
which yields $\H |\bk \rangle_{e/h} = \epsilon_{e/h}(\bk) |\bk \rangle_{e/h}$ (up to an irrelevant constant). The quasiparticle dispersion $\epsilon_{e/h}(\bk)$ given by
\begin{equation}
    \epsilon_{e/h}(\bk) = \epsilon_0(\bk) \pm \epsilon_F(\bk), \quad \epsilon_F(\bk) = \frac{1}{2A} \sum_\bq V_\bq |\lambda_\bq(\bk)|^2
\end{equation}
The interaction-generated dispersion $\epsilon_F(\bk)$ is the same as the Fock term in the Hartree-Fock Hamiltonian which yields a non-trivial band dispersion as long as the magnitude of the form factor $|\lambda_\bq(\bk)|$ is $\bk$-dependent. In the following, we will mainly focus on the electron bands and drop the $e$ subscript. 

We now consider a basis of states containing an electron and a spin flip which is obtained from the ground state ferromagnet by creating two electrons with spin up and a hole with spin down
\begin{equation}
    |\bk_{e1}, \bk_{e2}, \bk_{h} \rangle = c^\dagger_{\uparrow, \bk_{e1}} c^\dagger_{\uparrow, \bk_{e2}} c_{\downarrow, \bk_{h}}| \downarrow \rangle 
    \label{ke12h}
\end{equation}
The effective Hamiltonian in the two-electron/one-hole sector is defined as $H^{2e1h}_{\bk'_{e1}, \bk'_{e2}, \bk'_h; \bk_{e1}, \bk_{e2}, \bk_h} = \langle \bk'_{e1}, \bk'_{e2}, \bk'_h| \H | \bk_{e1}, \bk_{e2}, \bk_h \rangle$
 whose explicit form can be obtained from the commutation relations (\ref{Commutators}) and is provided in the supplemental material. The Hamiltonian $H^{2e1h}$ acts on the state $|\bk_{e1}, \bk_{e2}, \bk_{h} \rangle$ by shifting two of the three momenta $\bk_{e1}$, $\bk_{e2}$ and $\bk_h$ such that the total momentum $\bk = \bk_{e1} + \bk_{e2} - \bk_h$ is conserved. Thus, we can diagonalize the Hamiltonian for each total momentum sector separately and label the states by the two electronic momenta $\bk_{e1}$ and $\bk_{e2}$. Due to fermionic anticommutation relations $|\bk_{e1}, \bk_{e2} \rangle = -|\bk_{e2}, \bk_{e1} \rangle$, the states $|\bk_{e1}, \bk_{e2} \rangle$ only label $\frac{N(N-1)}{2}$ distinct states for a grid with $N$ points.

Notice that the Hilbert space spanned by the states (\ref{ke12h}) corresponds to states with definite $S_z = -\frac{N-3}{2}$ but with total spin $S = \frac{N-3}{2}$ or $\frac{N-1}{2}$. Since the Hamiltonian (\ref{Ham}) conserves the total spin, we can label the eigenstates of $H^{2e1h}$ by $S = \frac{N-1}{2}, \frac{N-3}{2}$. The explicit form of the total spin operator in the basis (\ref{ke12h}) is provided in supplemental material. Notice that the Hamiltonian $H^{2e1h}$ always has an eigenstates with total spin $S = \frac{N-1}{2}$ and energy $\epsilon(\bk)$ generated by acting with the spin raising operator, which commutes with $\H$, on the single particle excitation $|\bk \rangle$.

Our goal is to understand the energy competition between single particle excitations ($S = \frac{N-1}{2}$) and those dressed by a spin flip ($S = \frac{N-3}{2}$).
If the latter is lower in energy, this indicates the existence of a bound state of an electron and spin flip, a spin polaron. So far our discussion has been very general. Our model has as inputs the interaction $V_\bq$, the bare dispersion $\epsilon_0(\bk)$ and the form factors $\lambda_\bq(\bk)$ \footnote{Although $\epsilon_0(\bk)$ and $\lambda_\bq(\bk)$ are generally generated from the same microscopic model, we will think of them as independent which allows us for instance to take into account the effect of remote bands on the dispersion}. To study the competition between single-particle excitations and spin dressed excitations, we use a class of continuum models that contineously interpolates between the LLL and the narrow Chern bands of TBG. This class of models provide an excellent playground to study the formation of spin polarons since by allowing independent tuning of bandwidth, band topology and band geometry. 

{\em Twisted Bilayer Graphene Bands:} Before discussing our results, it is instructive to briefly review the continuum model for twisted bilayer graphene (TBG) \cite{dosSantos2007, dosSantos2012, Bistritzer2011} which consists of two Dirac Hamiltonians coupled through a Moir\'e potential. The latter has a matrix structure in the sublattice space and can be parametrized by two hopping parameters for intra- and inter-sublattice tunneling, denoted by $w_{\rm AA}$ and $w_{\rm AB}$ whose ratio has been estimated to be around $\kappa = \frac{w_{\rm AA}}{w_{\rm AB}} \approx 0.5 - 0.8$ \cite{Nam2017, Carr2018relax, Carr2019, Carr2020review, TBorNotTB}. A particularly interesting limit called the chiral limit corresponds to the case $\kappa = 0$ \cite{Tarnopolsky}. The wavefunctions of the model in this limit are sublattice-polarized with Chern number $\pm 1$ and has been shown to be equivalent to the lowest Landau level of a Dirac particle in an inhomogeneous periodic magnetic field \cite{LedwithFCI, LecNotes} providing a direct relation between this model and Landau level physics. Even away from the $\kappa = 0$ limit, we can define such sublattice-polarized basis where the bands have well-defined Chern numbers \cite{KIVCpaper}. This leads to a total of 4+4 bands with Chern numbers $+1$ and $-1$.

{\clr The results of this work apply to TBG under two assumptions. First, we neglect the coupling between Chern sectors which implies separate spin conservation in each sector that allows us to sharply distinguish the electron from the polaron (otherwise, the two can in principle tunnel into each other). Second, we assume only two active bands with $\SU(2)$ spin rotation. The remaining bands are assumed to be completely filled or empty and only influence the problem by changing the dispersion $\epsilon_0(\bk)$ through Hartree corrections as we will discuss later. At the end, we will discuss the validity of these assumptions.}

\emph{\bf Limit of flat quasiparticle dispersion}--- 
We will start by focusing on the limit of flat quasiparticle dispersion where the single particle dispersion and the interaction-generated dispersion exactly cancel for the electron band, $\epsilon_0(\bk) = -\epsilon_F(\bk)$.  Note that this is distinct from what is usually considered the flat-band limit where the bare dispersion vanishes $\epsilon_0(\bk) = 0$. The motivation for starting with this limit is that it allows us to isolate the effects of band geometry and topology from those of the dispersion, which will be added later. At the end, we will discuss the relevance of this limit to actual TBG. In general, we will define an effective magnetic field $B$ and magnetic length $l_B$ for any Chern $C$ band as 
\begin{equation}
    B = \frac{2\pi C}{A_{\rm BZ}}, \quad |B| = 2\pi l_B^2, 
\end{equation} 
Unless otherwise stated, we will be using the parameters for TBG at the magic angle $\theta = 1.0595^o$ with unscreened Coulomb interaction with relative permittivity $\epsilon = 10$.

\begin{figure*}
    \centering
    \includegraphics[width =  \textwidth]{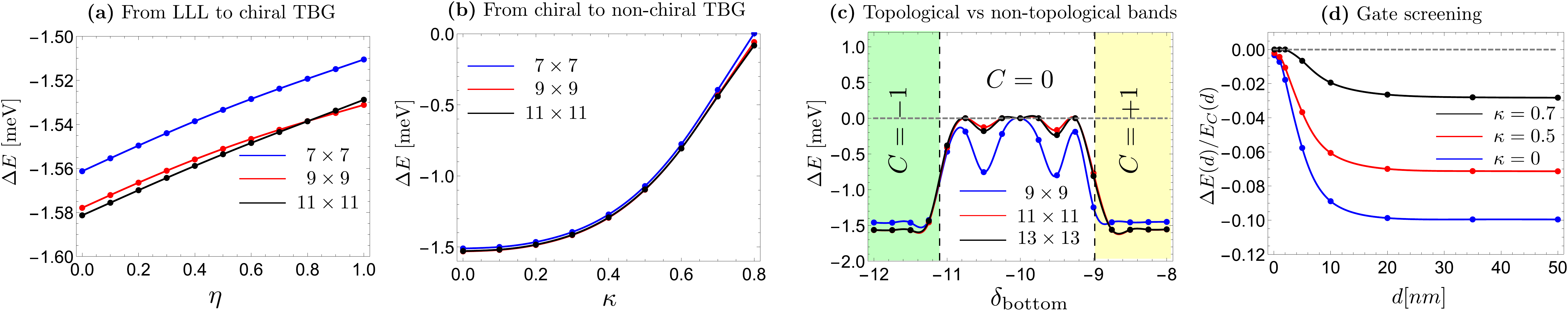}
    \caption{Binding energy of the polaron bound state for flat quasiparticle dispersion in different limits: (a) as we extrapolate between the LLL in uniform field and chiral TBG wavefunctions, (b) as a function of the chiral ratio $\kappa$, (c) as we tune the Chern number of the band by changing the bottom layer sublattice potential $\delta_{\rm bottom}$ for fixed top layer sublattice potential $\delta_{\rm top} = 10$ meV, and (d) as a function of the screening gate distance.}
    \label{fig:BindingEnergy}
\end{figure*}

{\em From LLL to TBG Chern Band:} Let us start with the simplest possible Chern band, the lowest Landau level. In order to compare this model with Chern bands defined in momentum space, we define a unit cell that contains a single flux quantum so that $A_{\rm BZ} = \frac{2\pi}{|B|}$. To connect this limit to chiral TBG, we introduce a periodic non-uniform component of the magnetic field whose average over the unit cell is zero and consider the LLL of a Dirac particle in such field. This introduces non-uniformity in the Berry curvature and allows us to continuously tune between the LLL and chiral TBG. More specifically, if the effective magnetic field for chiral TBG is $B_{\rm eff} = B_0 + B(\br)$ \cite{LedwithFCI, LecNotes}, we consider a Dirac particle in magnetic field $B_{\rm eff} = B_0 + \eta B(\br)$ where $\eta$ goes from 0 to 1 interpolating between the LLL and chiral TBG. We define $\Delta E$ to be the energy of the lowest $S = \frac{N-3}{2}$ eigenvalue of $H^{2e1h}$ relative to the energy of the lowest $S = \frac{N-1}{2}$. Note that in the thermodynamic limit, there is a continuum of $S = \frac{N-3}{2}$ states lying directly above the single particle $S = \frac{N-1}{2}$ state which implies that $\Delta E \leq 0$ \footnote{In the absence of a bound state, we numerically get a finite positive value for $\Delta E$ which scales to 0 with increasing system size. Throughout this work, we will set $\Delta E$ to zero in these cases}. $\Delta E$ is shown in Fig.~\ref{fig:BindingEnergy}a as a function of $\eta$. We can see clearly that its value is negative for all $\eta$ indicating the formation of a bound state of an electron and a spin-flip with binding energy around around $-1.5$ meV for our choice of parameters. Although changing $\eta$ introduces variations in the Berry curvature distribution, the energy of the bound state is essentially independent of $\eta$.

{\em Tuning Chiral Ratio:} Next, we introduce deviations from the chiral limit by considering non-zero value for the chiral ratio $\kappa$. Finite $\kappa$ is known to alter the geometric properties of the bands and cause the Berry curvature to be concentrated at $\Gamma$ \cite{ShangNematic,LedwithFCI, LecNotes}. For $\kappa \gtrapprox 0.7 - 0.8$, the Berry curvature is very close to a delta function at $\Gamma$ indicating the indicates the proximity of this value to a point where the remote bands touch the flat band and the band projection becomes invalid. In this limit, the Berry curvature can be gauged away  \cite{KhalafSoftMode} leading to the loss of the band's topological character. As we can see in Fig.~\ref{fig:BindingEnergy}b, the bound state persists for all values of $\kappa \lesssim 0.8$ but its binding energy start to approach zero as we approach the limit of very concentrated Berry curvature hinting at the topological origin of the bound state.

{\em Sublattice potential:} We can see the effect of topology more manifestly by considering a tuning parameter which alters the band topology by inducing a phase transition to a trivial Chern band. This is done by adding layer-dependent sublattice potential $\delta_{\rm top/bottom}$, which can be physically realized from aligned hBN \cite{Sharpe2019, YoungQAH}. As was shown in Ref.~\cite{Bultinck2019}, the sublattice polarized bands have vanishing Chern number when $\delta_{\rm top} + \delta_{\rm bottom}$ is close to 0, and finite Chern number $\pm 1$ otherwise. In Fig.~\ref{fig:BindingEnergy}c, we show $\Delta E$ as a function of $\delta_{\rm bottom}$ for fixed $\delta_{\rm top} = 10$ meV. We see that $\Delta E$ remains roughly constant on the topological side until we approach the transition where it rapidly increases till it vanishes in the non-topological side.

\emph{Gate screening:} So far we have been considering unscreened Coulomb interaction  relevant for TBG samples where the distance to the gate is much larger than the Moir\'e length scale. We will now consider the effect of gate screening by taking $V_\bq$ to be double-gate screened Coulomb interaction $V_\bq(d) = \frac{1}{2\epsilon \epsilon_0 |\bq|} \tanh q d$ with $d$ denoting the gate distance. Since changing $d$ also alters the overall energy scale, we will find it more useful to measure energy in terms of the scale $E_C(d) = \frac{1}{2A} \sum_\bq V_\bq(d) e^{-\frac{|B|}{2} \bq^2}$ which reduces to half the particle-hole gap for the LLL. The polaron binding energy is plotted as a function of $d$ in Fig.~\ref{fig:BindingEnergy}d  which shows how $\Delta E$ starts decreasing when the gate distance is around $10$ nm (roughly the Moir\'e scale) until it vanishes in the limit $d \rightarrow 0$. This is consistent with what is known about skyrmion energies which approaches the single-particle energy as the screening length is reduced. In fact, it was shown in Ref.~\cite{SkyrmionsWithoutSigmaModel} that for the LLL, the energy of skyrmions of \emph{any} size is the same as that of single-particle excitations for a delta potential, i.e. $d \rightarrow 0$.

\emph{\bf Topological electron-magnon coupling}--- To understand the existence of a bound state of an electron and a spin flip, it is instructive to rewrite the Hamiltonian $H^{2e1h}$ by labelling the Hilbert space of $S_z = -\frac{N-3}{2}$ in terms of an electron and a magnon excitation. The latter correspond to the eigenmodes of the Hamiltonian in the space of single spin flip operators
\begin{equation}
    a_{n,\bq}^\dagger = \sum_\bk c_{\bk,\uparrow}^\dagger c_{\bk + \bq, \downarrow} \phi_{n, \bq}(\bk), \quad \H a_{n,\bq}^\dagger |\downarrow \rangle = \xi_{n,\bq} a_{n,\bq}^\dagger |\downarrow \rangle
    \label{anq}
\end{equation}
where $\bq$ belongs to the first BZ. The operators $a_{n,\bq}^\dagger$ provide a complete $N$-dimensional orthonormal basis for spin flip operators which can be used to represent any spin flip operator $c_{\bk,\uparrow}^\dagger c_{\bk + \bq, \downarrow}$. The lowest energy state $n = 0$ corresponds to the Goldstone mode of the broken $\SU(2)$ spin symmetry whose dispersion satisfies $\xi_{0,\bq} \rightarrow 0$ as $\bq \rightarrow 0$. 

Using $a_{n,\bq}$, we can introduce the basis
\begin{equation}
    |\bk_0; \bq, n \rangle = c_{\bk_0 + \bq,\uparrow}^\dagger a^\dagger_{n,\bq} |\downarrow \rangle
    \label{Basisqn}
\end{equation}
which represents a state with total momentum $\bk_0$ (in our convention, $a^\dagger_{n,\bq}$ has momentum $-\bq$). Note that the state $|\bk_0; 0, 0 \rangle$ is a single-particle state with $S = \frac{N-1}{2}$ since $a_{0,\bq=0}^\dagger$ is simply the generator of a uniform spin rotation which increases $S_z$ of the ground state by $1$. The obvious advantage of this basis is that we expect low energy states to only have significant weight for small values of $n$ which in practice allows us to truncate the Hamiltonian for some $n$. However, one important subtlety about this basis is that it is not orthonormal. Instead, the overlap of states is given by
\begin{multline}
    g_{\bq,\bq'}^{nm}(\bk_0) = \langle \bk_0; \bq, n|\bk_0; \bq', m \rangle = \delta_{mn} \delta_{\bq,\bq'} \\ - \psi^*_{m \bq'}(\bk_0 + \bq) \psi_{n \bq}(\bk_0 + \bq')
\end{multline}
This operator is nothing but $1 - \hat F$ where $\hat F$ is the operator that exchanges two electrons. The basis (\ref{Basisqn}) contains $N^2$ states for a given $\bk_0$ which includes $\frac{N (N-1)}{2}$ fermionic (antisymmetric) states with $g(\bk_0)$ eigenvalues $2$ and $\frac{N (N+1)}{2}$ bosonic (symmetric) states with $g(\bk_0)$ eigenvalues $0$. Since the exchange operator $\hat F$ commutes with the Hamiltonian, we can obtain the physical Hilbert space (\ref{ke12h}) simply by restricting to the eigenstates of the Hamiltonian with $g(\bk_0)$ eigenvalue $2$.

In this basis, the Hamiltonian has the form
\begin{multline}
    \H |\bk_0; \bq, n \rangle = [\xi_{n,\bq} + \epsilon(\bk_0 + \bq)] |\bk_0; \bq, n \rangle \\ + \frac{1}{A} \sum_{\bq'} V_{\bq'} \lambda^*_{\bq'}(\bk_0 + \bq) C^{nm}_{\bq,\bq'} |\bk_0; \bq + \bq', m \rangle 
    \label{Heph}
\end{multline}
where $C^{nm}_{\bq,\bq'}$ are defined as
 \begin{equation}
        C^{nm}_{\bq,\bq'} = \sum_\bk \phi^*_{m,\bq + \bq'}(\bk) [\lambda_{\bq'}(\bk) \phi_{n,\bq}(\bk + \bq') - \phi_{n,\bq}(\bk) \lambda_{\bq'}(\bk + \bq) ]
        \label{Cnmqq}
    \end{equation}
The meaning of the different terms in the Hamiltonian above is transparent. The first two terms correspond to the magnon and the electron dispersion, respectively. These terms generally favor single particle excitations with $\bq = 0$ and $n = 0$ when $\bk_0$ is the dispersion minimum. 

The last term corresponds to the matrix elements of the interaction between electron-magnon states with magnon momenta $\bq$ and $\bq'$. If we focus on the lowest mode $n = m = 0$ and take the limit of small $\bq$ and $\bq'$, we find
\begin{equation}
    C^{00}_{\bq,\bq'} \approx i \bq \wedge \bq' B, \quad B = \frac{2\pi C}{A_{\rm BZ}}
    \label{C00qq}
\end{equation}
To see where this expression comes from, it is instructive to first consider the case of the LLL. One simplification we can do here is to unfold the BZ by extending $\bq$ beyond the first BZ and removing the index $n$. The coefficient $C_{\bq,\bq'}$ is then precisely the coefficient of the commutator of the GMP algebra $[\rho_\bq, a^\dagger_{\bq'}] = C_{\bq,\bq'} a^\dagger_{\bq + \bq'}$ \cite{GMP} which in this case is equal to $2i \sin \left(\frac{B}{2} \bq \wedge \bq'\right)$. For a general Chern band, the GMP algebra holds to linear order in $\bq$ and $\bq'$ \cite{Roy2014} with the prefactor given by $i B \bq \wedge \bq'$ which is precisely what we get in Eq.~\ref{C00qq}. A more detailed derivation of this result is given in supplemental material.

\begin{figure}
    \centering
    \includegraphics[width = 0.45 \textwidth]{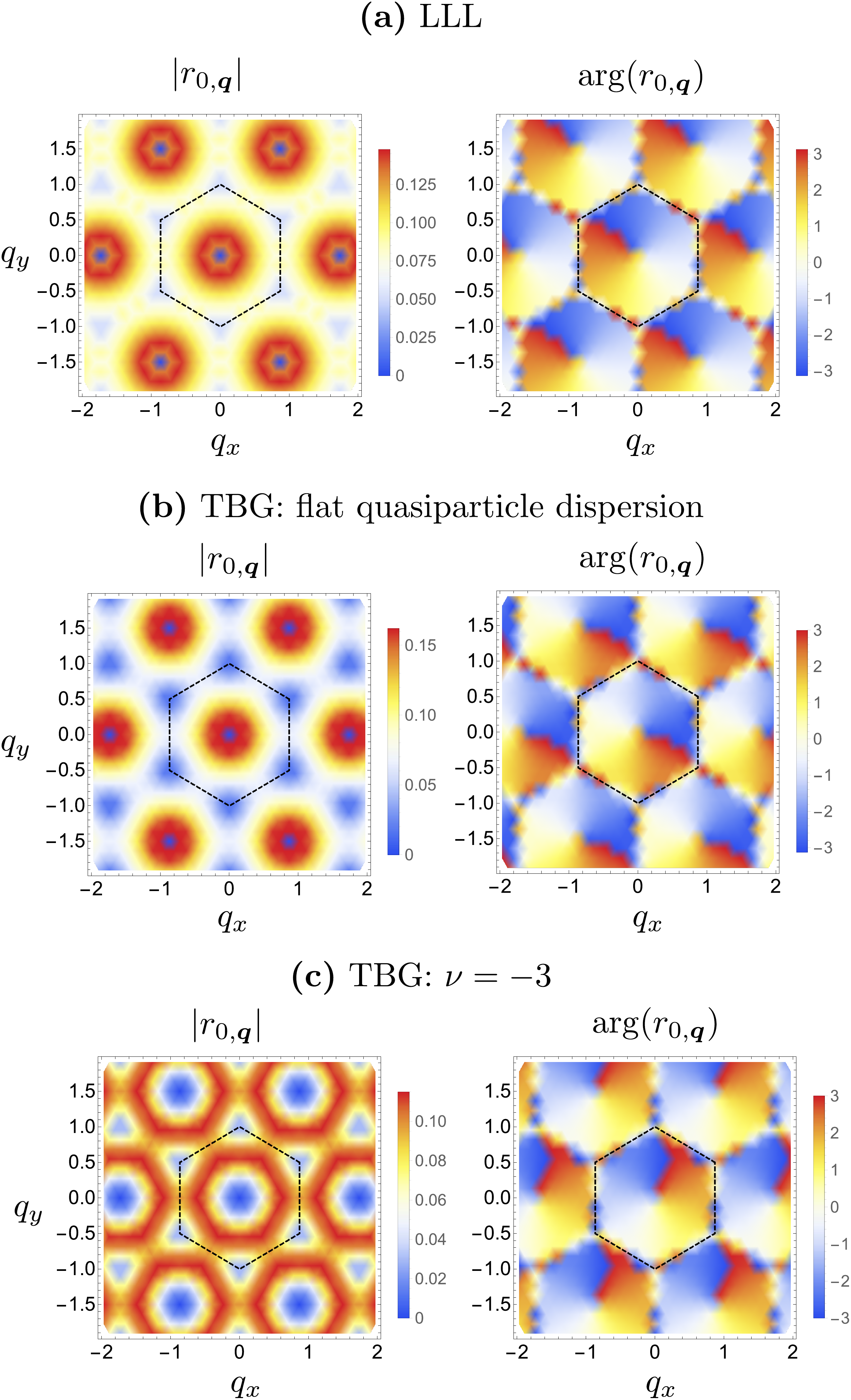}
    \caption{Wavefunctions of the bound state as a function of the magnon momentum $\bq$ (cf.~Eq.~\ref{Psi}) for {\bf (a)} the LLL, {\bf (b)} TBG with flat quasiparticle dispersion, and {\bf (c)} for electron (hole) doping the $\nu = -3$ ($\nu = +3$) TBG bands. We used $\kappa = 0.55$ and an $11 \times 11$ grid.}
    \label{fig:Wavefunctions}
\end{figure}

Let us now we write a general eigenstate of (\ref{Heph}) as
\begin{equation}
    |\Psi \rangle = \sum_{n,\bq} r_{n,\bq} |\bk_0; \bq, n \rangle
    \label{Psi}
\end{equation}
Focusing on the $n = 0$ component in the small $\bq$ limit, we see from (\ref{C00qq}) that the magnitude of the last term in the Hamiltonian (\ref{Heph}) is maximized when connecting states $|\bk_0; \bq, n \rangle$ and $|\bk_0; \bq', n \rangle$ with $\bq$ and $\bq'$ orthogonal, i.e. if they are related by a $\pi/2$ rotation. Furthermore, due to the factor of $i$ in (\ref{C00qq}), we can make this term negative by choosing $r_{0,\bq}$ to change its phase by $\pi/2$ upon rotating $\bq$ by $\pi/2$. Thus, we can minimize this term by choosing $r_{0,\bq} \propto e^{i \arg(q_x + i q_y)}$. In addition, since this term vanishes when $\bq$ or $\bq'$ vanish, the magnitude of $r_{0,\bq}$ should not vanish too quickly with $\bq$. We can see this by expanding the first two terms in (\ref{Heph}) at small momenta $\xi_{0,\bq} \sim l_B^2 \rho \bq^2$ and $\epsilon(\bk_0 + \bq) \sim \frac{l_B^2}{m_{\rm eff}} \bq^2$. Then, if we assume that $r_{0,\bq}$ decays for momenta larger than some cutoff $\Lambda$, we find that the first two terms in the Hamiltonian give a positive energy contribution of order $l_B^2 (\rho + m_{\rm eff}^{-1}) \Lambda^2$ whereas the last term gives a negative contribution of order $E_C l_B^3 \Lambda^3$. Thus, a bound state has to have a finite extent  in momentum of at least $\Lambda \sim \frac{\rho + m^{-1}}{E_C l_B}$. This is verified by plotting $r_{0,\bq}$ for both the LLL and chiral TBG in Fig.~\ref{fig:Wavefunctions}. We see that $|r_{0,\bq}|$ decays in $\bq$ within the first BZ and that $\arg r_{0,\bq}$ winds by $2\pi$ around the $\Gamma$ point. For the LLL case, due to continuous magnetic translation, we can unfold the BZ and write the variational state $r_\bq = e^{-\frac{\xi}{2} |\bq| + i \arg (q_x + i q_y)}$ whose overlap with the numerically obtained solution exceeds 99\% for appropriately chosen $\xi$ (see supplemental material for details). 

\emph{\bf Finite quasiparticle dispersion}--- Let us now consider the limit of dispersive quasiparticle bands. Motivated by the energetics of TBG bands, we will choose $\epsilon_0(\bk) = \nu \epsilon_H(\bk)$ where $\epsilon_H(\bk) = \frac{1}{A} \sum_\bG V_\bG \lambda_\bG(\bk) \sum_{\bk'} \lambda_{-\bG}(\bk')$ is the Hartree potential \cite{XieMacdonald, ShangNematic, Guinea2018, TBGV, PierceCDW, VafekBernevig}. This form of the dispersion directly allows us to compare our results to TBG since the quasiparticle dispersion 
\begin{equation}
    \epsilon_\nu(\bk) = \nu \epsilon_H(\bk) + \epsilon_F(\bk)
    \label{enuk}
\end{equation}
approximately describes the dispersion of electron (hole) bands for a correlated insulator at integer filling $\nu$ ($-\nu$) \cite{XieMacdonald, ShangNematic, GuineaHF, TBGV, PierceCDW, VafekBernevig}. Although the expression (\ref{enuk}) is only valid for integer $\nu$, we will choose to take $\nu$ to be a continuous variable which allows us to continuously tune the band dispersion. It also makes our results less sensitive to uncertainty in model parameters. For instance, while our model is particle-hole symmetric, we can phenomenologically incorporate partice-hole asymmetry by shifting $\nu$ which approximately captures the effects discussed in Ref.~\cite{MacdonadPH}. $\epsilon_H(\bk)$ is characterized by a dip at the $\Gamma$ point and thus have a qualitatively similar shape to the Fock term $\epsilon_F(\bk)$ up to overall scaling. Thus for $\nu > 0$, the two terms add, leading to a large bandwidth while for $\nu < 0$ they subtract leading to a reduced bandwidth \cite{XieMacdonald, TBGV, PierceCDW, VafekBernevig}. The minimum bandwidth is realized for $\nu \approx -1$ to $-1.5$ depending on the value of $\kappa$ as shown by the dashed line in Fig.~\ref{fig:Dispersion}a. The details of the dispersion are reviewed in the supplemental material.

\begin{figure}
    \centering
    \includegraphics[width = 0.45 \textwidth]{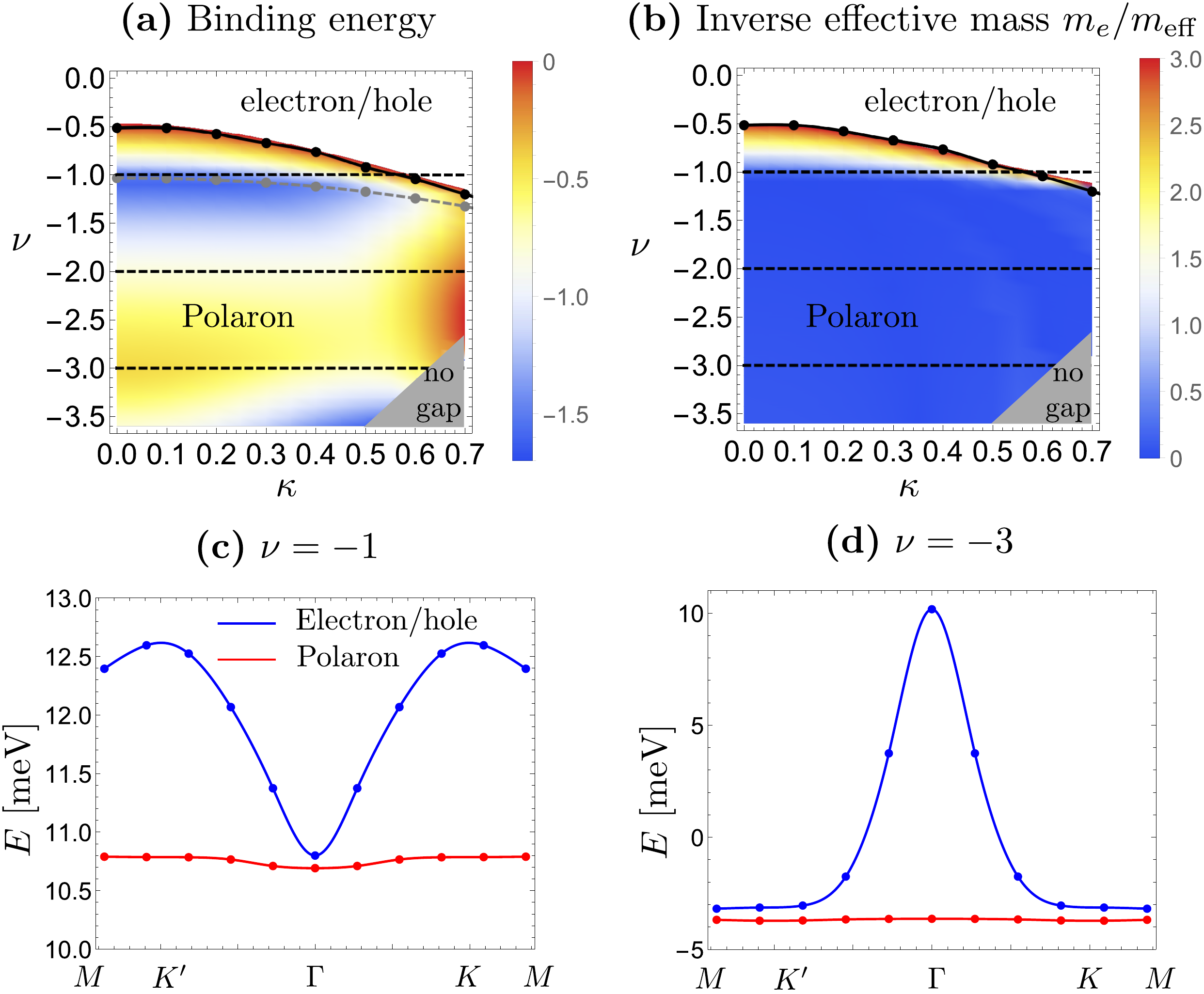}
    \caption{{\bf (a)} Binding energy as a function of ``filling" $\nu$ and chiral ratio $\kappa$. The dashed lines indicate the energies for electron (hole) doping the insulator at filling $\nu$ ($-\nu$), i.e. doping towards neutrality. $\nu > 0$ (doping away from neutrality) is not shown since no bound state is found. The solid black line indicates the boundary where the bound state is lost, the dashed gray line corresponds to the minimum bandwidth, and the gray shaded area is where minimum of the conduction band is lower in energy than the top of the valence band and the assumptions of our analysis become invalid. {\bf (b)} The corresponding inverse effective mass at the bottom of the band. The phase boundary is well approximated by the value $m_e/m_{\rm eff} \approx 3$. {\bf (c)} and {\bf (d)} are the spectra for electron doping the $\nu = -1$ and $\nu = -3$ insulators, respectively (or alternatively hole doping at $\nu = +1$ and $\nu = +3$). For {\bf (c)} and {\bf (d)}, we used $\kappa = 0.55$.}
    \label{fig:Dispersion}
\end{figure}

We can investigate the existence of a bound state as a function of dispersion parameterized by $\nu$. We find that there is a critical value $\nu_c$, indicated by the solid line in Fig.~\ref{fig:Dispersion}a, such that a bound state exists iff $\nu < \nu_c$. We see that this value is always negative and ranges from around $-0.5$ in the chiral limit to around $-1.2$ at $\kappa = 0.7$. The implication for TBG is that, generally, polaron formation is favored on electron (hole) doping the $\nu < 0$ ($\nu > 0$) insulators, i.e.  doping \emph{towards} neutrality. On doping away from neutrality, we always find the single particle excitations to be the lower in energy. We note that for $\nu$ sufficiently large and negative, the Hartree term dominates and we get a peak rather than a dip at $\Gamma$. In this case, we always find a polaron bound state even though the bandwidth can be quite large. This rather surprising results can be explained by examining the wavefunction of the $\Gamma$ polaron in Fig.~\ref{fig:Wavefunctions}c. We see that, compared to the case of flat quasiparticle dispersion, the wavefunction has suppressed weight at small $\bq$ enabling it to avoid the energetically costly region around $\Gamma$.

This suggests that the formation of a bound state is mostly sensitive to the effective mass at the bottom of the band, which is relatively large at the band minimum for $\nu$ large and negative, rather than the total bandwidth. The effective mass has the added advantage of being an experimentally accessible quantity, e.g. from quantum oscillations \cite{PabloMott, PabloSC}, allowing us to make phenomenological comparison with experiments that are not tied to theory parameters. In Fig.~\ref{fig:Dispersion}b, we show the effective mass for different values ($\nu$, $\kappa$) and compare it to the phase diagram in Fig.~\ref{fig:Dispersion}a. We find that, quite remarkably, the phase boundary where the polaron bound state is lost can be described very well by the expression $m_e/m_{\rm eff} \approx 3$ as shown in Fig.~\ref{fig:Dispersion}b. 
{\clr This value is not far off the experimentally extracted value $m_{\rm eff}/m_e \approx 0.2 - 0.3$ from quantum oscillations for small hole doping of the $\nu = -2$ state where superconductivity was first seen \cite{PabloSC}. This suggests that the experimental regime for superconductivity can be close to the phase boundary where a bound state would form even when the lowest charged excitation are single-particle like.}

Finally, we investigate the dispersion of the spin polaron state in the limit when it is the lowest energy excitation. We consider two cases: (i) electron (hole) doping the $\nu = -1$ ($\nu = 1$) insulator where the dispersion {\em minimum} is at $\Gamma$ and (ii) electron (hole) doping the $\nu = -3$ ($\nu = 13$) insulator where the dispersion {\em maximum} is at $\Gamma$. The resulting dispersion is shown in Figs.~\ref{fig:Dispersion}c and d. We see that the polarons have much flatter bands with significantly larger effective mass that can be as large as 30 times the electron effective mass. %{\clr The wavefunctions of the $\Gamma$ polaron bound states for $\nu = -1$, where the dispersion minimum is at $\Gamma$, look very similar to the flat quasiparticle dispersion limit shown in Fig.~\ref{fig:Wavefunctions}b. On the other hand, for $\nu = -3$, where the band maximum is at $\Gamma$, the wavefunctions for the $\Gamma$ polaron have very small wieght at small $\bq$ to avoid the energetically costly region around the $\Gamma$ point as shown in Fig.~\ref{fig:Wavefunctions}c}.

\emph{\bf Implications for TBG}--- Before discussing potential implications for our results, let us point out a few caveats. First, the lowest energy charged excitations we obtain here are just among single particle states and those dressed by a single spin flip. This means that it is always possible that states with more spin flips corresponding to larger skyrmions are lower in energy. This is known to be the case for the LLL \cite{Sondhi, MoonMori} and will likely hold here too when the bands are sufficiently flat. Furthermore, in some instances the energy as a function of number of spin flips may not be monotonic, and larger skyrmions can be favored even though the spin polaron is not.  

Second, our analysis focused on a single Chern sector. While we took into account Hartree-Fock interaction-generated dispersion which is mainly diagonal in the Chern index, we have neglected the part of the dispersion connecting different Chern sectors which have two main sources: (i) non-interacting dispersion which is sensitive to the twist angle and other parameters such as strain \cite{BiFuStrain, Parker}, and (ii) dispersion generated from the subtraction scheme employed when projecting-out the remote bands to avoid double counting \cite{XieMacdonald,RepellinFerromagnetism, ShangNematic, KIVCpaper}. Both terms take the form of tunneling between opposite Chern sectors which influences the physics in several ways. First, the electron and polaron are no longer distinguished by their spin quantum number since the spin is no longer conserved in a given Chern sector and thus they can tunnel into each other. Second, such tunneling perturbatively generates a dispersion that counters the Fock dispersion and reduces the effective mass at $\Gamma$. Third, it generates an antiferromagnetic `superexchange' coupling \cite{KIVCpaper, SkPaper, KhalafSoftMode} between the Chern sectors which alters the energetics of magnons and plays a crucial role in the skyrmion pairing scenario \cite{SkPaper, chatterjee2020skyrmion}. Thus, we expect these terms to favor polaron {\em pairs} (bipolarons) over  single electrons or electron pairs. We leave a detailed analysis of the effect of these terms to future works.

{\clr Finally, we have only focused on $\SU(2)$ polarons which assume there are only two active bands while the remaining bands are frozen. This approach can be phenomenologically justified by the observation of flavor polarization `cascade' features at relatively high temperatures \cite{CascadeShahal, CascadeYazdani} suggesting these polarized flavor degrees of freedom becomes frozen at low temperature. We note however that this is not generally true for the Hartree-Fock dispersion where the bands for the fully filled/empty flavors are not energetically separated from the remaining bands. This allows for the possibility of more complicated $\SU(4)$ skyrmion/polaron textures.}

Our findings suggests the following picture for charge excitations in TBG: (i) On doping a correlated insulator at integer $\nu \neq 0$ towards charge neutrality, charge enters the system as polarons or large skyrmions. This is consistent with the observed absence of quantum oscillations for this doping range \cite{PabloMott, PabloSC, Dean-Young} and also explains the rapid loss of flavor (spin/valley) polarization (cascade transition) \cite{CascadeShahal, CascadeYazdani} with doping. Combined with the observations that polarons are disfavored by reducing the screening length, this leads to the prediction that the cascade features should become weaker as the screening length to the gate is reduced \cite{YoungScreening, EfetovScreening}. (ii) On doping away from neutrality, charge likely enters the system as single-particle excitations consistent with the observation of Landau fans. Finally, Although we have focused on charge $e$ excitations, let us make a few observations about pairing, i.e. the charge $2e$ excitations. Pairing between stable spin-polarons, the analog of the skyrmion pairing mechanism proposed by the authors in Ref.~\cite{SkPaper} (see also Refs.~\cite{chatterjee2020skyrmion, ChristosWZW}) can be naturally associated with doping towards neutrality where superconductivity has been observed in some samples \cite{PabloSC, Efetov, EfetovScreening}. On the other hand, on doping away from neutrality (where superconductivity is seemingly more ubiquitous), spin polarons can remain relevant as finite energy long-lived excitations whose pairing correlations can be induced to the electrons, even when they are not the lowest charge excitations. {\clr This suggests a BEC-to-BCS scenario with increasing dispersion where the bound state is lost while superconductivity persists \cite{LeggettBECBCS, NozieresBECBCS, GurarieBECBCS}.} The latter limit may be related to the recently proposed scenario where pairing takes place by the exchange of pseudospin fluctuations \cite{MacdonaldPseudospin}. 
A detailed theory of spin-polaron  pairing will be the topic of a future work.

\emph{\bf Conclusion}--- {\clr In summary, we have identified a general tendency for the formation of a polaron bound state between an electron and a spin flip in a Chern band that is purely topological in origin. We have studied the formation of such bound states over a wide range of parameters for the Chern bands of twisted bilayer graphene. This lead us to identify experimental parameter range where spin polarons are formed and discuss their possible experimental consequences. Our results highlight the surprising fact  that although the ground state is well approximated by a Slater determinant,  a description in terms of electron like single-particle excitations, whether approximate or exact, is insufficient to describe the charge physics in Chern bands. Furthermore, our analysis serves as a bridge between real space skyrmion textures and single-particle excitations.

\emph{Note}--- We would like to point out a parallel work \cite{YvesSkyrmions} which gives an extensive report  of the energetics of TBG skyrmions, both charge `$e$' and charge `$2e$', using variational Hartree-Fock. The results of that work, which is suited to study the limit of large skyrmions, is complementary to our momentum space approach.

\emph{Acknowledgements:} E.~K. acknowledges stimulating discussions with Mike Zaletel, Herb Fertig, Patrick Ledwith and Dan Parker. {\clr We are grateful for Yves Kwan, Nick Bultinck and Sid Parameswaran for informing us about their concurrent study \cite{YvesSkyrmions}}. A.~V., E.~K. were funded by a Simons Investigator grant (AV) and by the Simons Collaboration on Ultra-Quantum Matter, which is a grant from the Simons Foundation (618615, A.V.).

\bibliographystyle{unsrt}
\bibliography{refs}

\pagebreak
\widetext
\begin{center}
\textbf{\large SUPPLEMENTAL MATERIAL:\\ Charge excitations in Chern ferromagnets}
\end{center}
%%%%%%%%%% Merge with supplemental materials %%%%%%%%%%
%%%%%%%%%% Prefix a "S" to all equations, figures, tables and reset the counter %%%%%%%%%%
\setcounter{equation}{0}
\setcounter{figure}{0}
\setcounter{table}{0}
\setcounter{page}{1}
\makeatletter
\renewcommand{\theequation}{S\arabic{equation}}
\renewcommand{\thefigure}{S\arabic{figure}}
\renewcommand{\bibnumfmt}[1]{[S#1]}
\section{Explicit form of the Hamiltonian and total spin operators}
The explicit form of the Hamiltonian $\H^{2e1h}$ can be easily obtained from the action of the Hamiltonian $\H$, Eq.~\ref{Ham}, on $|\bk_{e1}, \bk_{e2}, \bk_{h} \rangle$, defined in Eq.~\ref{ke12h}, using the commutation relations (\ref{Commutators}) leading to
\begin{multline}
    \H |\bk_{e1}, \bk_{e2}, \bk_{h} \rangle = [\epsilon_e(\bk_{e1}) + \epsilon_e(\bk_{e2}) + \epsilon_h(\bk_h)] |\bk_{e1}, \bk_{e2}, \bk_{h} \rangle + \frac{1}{A} \sum_\bq V_\bq \left\{ \lambda_\bq^*(\bk_{e1}) \lambda_{-\bq}^*(\bk_{e2}) |\bk_{e1} + \bq, \bk_{e2} - \bq, \bk_{h} \rangle \right. \\ \left. -  \lambda_\bq^*(\bk_{e1}) \lambda_{\bq}(\bk_h) |\bk_{e1} + \bq, \bk_{e2} , \bk_{h} + \bq \rangle -  \lambda_\bq^*(\bk_{e2}) \lambda_{\bq}(\bk_h) |\bk_{e1}, \bk_{e2} + \bq , \bk_{h} + \bq \rangle \right\}
\end{multline}

To express the total spin operator in the basis $|\bk_{e1}, \bk_{e2}, \bk_h \rangle$, we start the standard expression
\begin{gather}
    S^2 = S_x^2 + S_y^2 + S_z^2, \\ S_x = \frac{1}{2} \sum_\bk (c_{\bk,\uparrow}^\dagger c_{\bk, \downarrow} + c_{\bk,\downarrow}^\dagger c_{\bk, \uparrow}), \quad S_y = - \frac{i}{2} \sum_\bk (c_{\bk,\uparrow}^\dagger c_{\bk, \downarrow} - c_{\bk,\downarrow}^\dagger c_{\bk, \uparrow}), \quad S_z = \frac{1}{2} \sum_\bk (c_{\bk,\uparrow}^\dagger c_{\bk, \uparrow} - c_{\bk,\downarrow}^\dagger c_{\bk, \downarrow})
\end{gather}
The state $|\bk_{e1}, \bk_{e2}, \bk_h \rangle = c_{\bk_{e1},\uparrow}^\dagger c_{\bk_{e2},\uparrow}^\dagger c_{\bk_h,\downarrow} |\downarrow \rangle$ is an $S_z$ eigenstate with eigenvalue $-\frac{N-3}{2}$. The action of $S_x$ and $S_y$ can be obtained using the commutations relations
\begin{gather}
    [S_x, c_{\bk,\uparrow}^\dagger] = \frac{1}{2} c_{\bk,\downarrow}^\dagger, \qquad [c_{\bk,\downarrow}^\dagger, S_x] = -\frac{1}{2} c_{\bk,\uparrow}^\dagger, \qquad [S_x, c_{\bk,\downarrow}] = -\frac{1}{2} c_{\bk,\uparrow}, \qquad [c_{\bk,\uparrow}, S_x] = \frac{1}{2} c_{\bk,\downarrow} \\
    [S_y, c_{\bk,\uparrow}^\dagger] = \frac{i}{2} c_{\bk,\downarrow}^\dagger, \qquad [c_{\bk,\downarrow}^\dagger, S_y] = \frac{i}{2} c_{\bk,\uparrow}^\dagger, \qquad [S_y, c_{\bk,\downarrow}] = -\frac{i}{2} c_{\bk,\uparrow}, \qquad [c_{\bk,\uparrow}, S_y] = - \frac{i}{2} c_{\bk,\downarrow}
\end{gather}
which leads after straightforward but tedious calculation to
\begin{gather}
    S_x^2 |\bk_{e1}, \bk_{e2}, \bk_h \rangle = \frac{1}{4} \left[-3 |\bk_{e1}, \bk_{e2}, \bk_h \rangle + 2 \sum_\bk (-\delta_{\bk_{e1},\bk_h} |\bk_{e2}, \bk, \bk \rangle + \delta_{\bk_{e2},\bk_h} |\bk_{e1}, \bk, \bk \rangle ) \right] + c_{\bk_{e1},\uparrow}^\dagger c_{\bk_{e2},\uparrow}^\dagger c_{\bk_h,\downarrow} S_x^2 |\downarrow \rangle \\
    S_y^2 |\bk_{e1}, \bk_{e2}, \bk_h \rangle = -\frac{1}{4} \left[3 |\bk_{e1}, \bk_{e2}, \bk_h \rangle + 2 \sum_\bk (\delta_{\bk_{e1},\bk_h} |\bk_{e2}, \bk, \bk \rangle - \delta_{\bk_{e2},\bk_h} |\bk_{e1}, \bk, \bk \rangle ) \right] + c_{\bk_{e1},\uparrow}^\dagger c_{\bk_{e2},\uparrow}^\dagger c_{\bk_h,\downarrow} S_y^2 |\downarrow \rangle
\end{gather}
Which leads to
\begin{gather}
    S^2 |\bk_{e1}, \bk_{e2}, \bk_h \rangle = \left(\frac{N-1}{2} \right) \left( \frac{N-3}{2} \right) |\bk_{e1}, \bk_{e2}, \bk_h \rangle + \hat M |\bk_{e1}, \bk_{e2}, \bk_h \rangle, \\  \hat M |\bk_{e1}, \bk_{e2}, \bk_h \rangle = \sum_\bk (-\delta_{\bk_{e1},\bk_h} |\bk_{e2}, \bk, \bk \rangle + \delta_{\bk_{e2},\bk_h} |\bk_{e1}, \bk, \bk \rangle )
\end{gather}
It is easy to verify that the operator $\hat M$ defined on the second line satisfies $\hat M^2 = (N-1) \hat M$, hence its eigenvalues are 0 and $N-1$. The former yields $S^2$ eigenvalue $\left(\frac{N-1}{2} \right) \left( \frac{N-3}{2} \right)$ which corresponds to a total spin $S = \frac{N-3}{2}$ whereas the latter yields $S^2$ eigenvalue $\left(\frac{N-1}{2} \right) \left( \frac{N+1}{2} \right)$ which corresponds to a total spin $S = \frac{N-1}{2}$.

\section{Derivation of the topological electron-magnon coupling at small momenta}
Our purpose in this section is to derive the form of the electron-magnon coupling at small momenta, Eq.~\ref{C00qq}. The magnon  creation operator is defined in Eq.~\ref{anq}, repeated here for completeness
\begin{equation}
    a_{n,\bq}^\dagger = \sum_\bk c_{\uparrow,\bk}^\dagger c_{\downarrow,\bk + \bq} \phi_{n,\bq}(\bk)
\end{equation}
where $\phi_{n,\bq}(\bk)$ is the complete orthonornal set of eigenfunctions of the soft mode Hamiltonian defined as
\begin{equation}
    \H_\bq(\bk',\bk) = \langle  c_{\downarrow,\bk' + \bq }^\dagger c_{\uparrow,\bk'} \H_V  c^\dagger_{\uparrow,\bk} c_{\downarrow,\bk + \bq }\rangle, \qquad \sum_{\bk'} \H_\bq(\bk, \bk') \phi_{n,\bq}(\bk') = \xi_{n,\bq} \phi_{n,\bq}(\bk)
\end{equation}
We notice that gauge invariance requires that $\phi_{n,\bq}(\bk)$ transforms the same way as $\lambda_\bq(\bk)$ under gauge transformations. That is, under $c_{\bk,\sigma} \mapsto c_{\bk,\sigma} e^{i \theta_k}$, $\phi_{n,\bq}(\bk) \mapsto e^{-i [\theta_{\bk + \bq} - \theta_\bk]} \phi_{n,\bq}(\bk)$. This means we can define a gauge invariant $\tilde \phi_{n,\bq}(\bk)$ via
\begin{equation}
    \phi_{n,\bq}(\bk) = \tilde \lambda_\bq(\bk) \tilde \phi_{n,\bq}(\bk), \qquad \tilde \lambda_\bq(\bk) = \frac{\lambda_\bq(\bk)}{|\lambda_\bq(\bk)|}
    \label{tphi}
\end{equation}
where we used the phase of the form factor $\tilde \lambda_\bq(\bk)$ rather than its full value to maintain the normalization of the wavefunctions. It is easy to show  that in the limit $\bq \rightarrow 0$, $\phi_{0,\bq}(\bk) \rightarrow \frac{1}{\sqrt{N}}$. This is nothing but the statement that the Goldstone mode in the limit of long wavelength reduces to the spin raising operator. Thus, we can write
\begin{equation}
    \tilde \phi_{0,\bq}(\bk) \approx \frac{1}{\sqrt{N}} [1 + i \bq \cdot \bv(\bk) + O(\bq^2)]
    \label{tphiq}
\end{equation}
Crucially, we can show that the Hamiltonian $\tilde \H_\bq(\bk',\bk) = \tilde \lambda^*_\bq(\bk) \H_\bq(\bk',\bk) \tilde \lambda_\bq(\bk)$ is periodic and smooth in $\bk$, and so is $\tilde \phi_{n,\bq}(\bk)$ and $\bv(\bk)$. This can be seen by writing the transformed Hamiltonian $\tilde H_\bq(\bk,\bk') = \tilde \lambda_\bq^*(\bk) H_\bq(\bk,\bk') \tilde \lambda_\bq(\bk')$ and noting that it only depends on gauge invariant combinations of $\tilde \lambda_\bq(\bk)$ which can be written in terms of the Berry curvature which is periodic and smooth in $\bk$. Substituting (\ref{tphi}) and (\ref{tphiq}) in Eq.~\ref{Cnmqq} and using the small $\bq$ expansion of the form factor $\lambda_\bq(\bk) \approx 1 + i \bq \cdot \bA(\bk) + O(\bq^2)$ yields
 \begin{align}
      C^{00}_{\bq,\bq'} &= \frac{i}{N} \sum_\bk  \left\{q'_\mu [A^\mu(\bk) - A^\mu(\bk + \bq)] + q_\mu [A^\mu(\bk + \bq') - A^\mu(\bk)] + q_\mu [v^\mu( \bk + \bq') - v^\mu(\bk)] \right\} \nonumber \\
      &= i \bq_\mu \bq'_\nu \int \frac{d^2\bk}{A_{\rm BZ}} [\partial_\mu A_\nu(\bk) - \partial_\nu A_\mu(\bk)] = i \bq \wedge \bq' \int \frac{d^2\bk}{A_{\rm BZ}} \Omega(\bk) = i \frac{2\pi C}{A_{\rm BZ}} \bq \wedge \bq' 
 \end{align}
 On going from the first to the second line, we used the periodicity of $\bv(\bk)$ to shift the momentum summation leading to $\sum_\bk v^\mu( \bk + \bq') - \sum_\bk v^\mu(\bk) = 0$ (notice that this does not work for $A_\mu(\bk)$ which cannot be periodic in a band with finite Chern number). In the last equality, we used the definition of the Chern number $\int d^2 \bk \; \Omega(\bk) = 2\pi C$.
% Under this transformation, the Hamiltonian becomes
% \begin{equation}
%     \tilde \H_\bq(\bk',\bk) = \tilde \lambda^*_\bq(\bk) \H_\bq(\bk',\bk) \tilde \lambda_\bq(\bk) =  \frac{\delta_{\bk,\bk'}}{2A} \sum_{\bq'} V_{\bq'} (|\lambda_{\bq'}(\bk)|^2 + |\lambda_{\bq'}(\bk + \bq)|^2) - \frac{1}{A} \sum_\bG V_{\bk' - \bk + \bG} \lambda_{\bk' - \bk + \bG}(\bk + \bq) \lambda^*_{\bk' - \bk + \bG}(\bk)
% \end{equation}

\section{Spin polarons in Lowest Landau level}
In this appendix, we consider in some detail the formation of the spin polaron bound state in the LLL. The problem simplifies in this case due to contineous magnetic translation symmetry. The form factors for the LLL are given by
\begin{equation}
    \lambda_\bq(\bk) = \langle u_\bk| u_{\bk + \bq} \rangle = e^{-\frac{|B|}{4} q^2 - i \frac{B}{2} \bq \wedge \bk}
\end{equation}
where as in the main text, we take a unit cell which encloses one magnetic flux such that $|B| = \frac{2\pi}{A_{\rm BZ}}$. We can now use the properties of the GMP algebra \cite{GMP} to understand the formation the bound states.

In the following, we will follow the notation of Ref.~\cite{SkPaper} where $\hat A = \sum_{\alpha,\beta} c^\dagger_\alpha A_{\alpha \beta} c_\beta$ and introducing the matrix $[\lambda_q]_{\bk,\bk'} = \lambda_\bq(\bk) \delta_{\bk',\bk + \bq}$ such that $\rho_\bq = \hat \lambda_\bq$. We will also use the relation \cite{SkPaper}
\begin{equation}
    [\hat A, \hat B] = \widehat{[A, B]}
\end{equation}
We can see that the GMP algebra \cite{GMP} follows from the matrix commutation relation
\begin{align}
    [\lambda_\bq, \lambda_{\bq'}]_{\bk,\bk'} &= \delta_{\bk',\bk + \bq + \bq'} [\lambda_\bq(\bk) \lambda_{\bq'}(\bk + \bq) - \lambda_{\bq'}(\bk) \lambda_{\bq}(\bk + \bq')] =  2 i \sin \frac{B}{2} \bq \wedge \bq' \, e^{-\frac{B}{4} (\bq^2 + \bq'^2) - i \frac{B}{2} (\bq + \bq') \wedge \bk} \delta_{\bk', \bk + \bq + \bq'} \nonumber \\
    &= 2 i \sin \left(\frac{B}{2} \bq \wedge \bq'\right) e^{\frac{B}{2} \bq \cdot \bq'} [\lambda_{\bq + \bq'}]_{\bk,\bk'}
\end{align}

For this choice of of form factors, we need to make a specific gauge choice for the operators $c_\bk$ such that $\rho_\bq$ does not change under shifting $\bk$ in the summation by a reciprocal lattice vector $\bG$. This is realized by taking
\begin{equation}
    c_\bk = c_{[\bk]} e^{i \phi_\bk}, \qquad \phi_\bk =  - \frac{B}{2} \{\bk \} \wedge [\bk] - \frac{B}{4} \{\bk \} \wedge \{\bar \bk \}
\end{equation}
where we decomposed $\bk$ into $[\bk]$ which lies in the first BZ and $\{\bk\} = \bk - [\bk]$ which is a reciprocal lattice vector. We also introduced the bar notation such that for a vector $\bk = k_1 \bb_1 + k_2 \bb_2$, where $\bb_1$ are the momentum space basis vectors, $\bar \bk = k_1 \bb_1 - k_2 \bb_2$. To see how this makes $\rho_\bq$ invariant under shifts in $\bk$, we note the following identities. Assuming $\bb_1 \wedge \bb_2 = + A_{\rm BZ}$, we can write
\begin{gather}
    \bn = n_1 \bb_1 + b_2 \bb_2, \quad \bm = m_1 \bb_1 + m_2 \bb_2, \\
    \bn \wedge \bar \bm = - A_{\rm BZ} (n_1 m_2 + n_2 m_2) = \bm \wedge \bar \bn, \qquad \bn \wedge \bar \bn = - 2 A_{\rm BZ} n_2 n_2
\end{gather}
Thus, we get
\begin{equation}
    c_{\bk + \bG} = c_\bk e^{-i \frac{B}{2} \bG \wedge [\bk] - i \frac{B}{2} \{\bk \} \wedge \bar \bG - i \frac{B}{4} \bG \wedge \bar \bG }, 
\end{equation}
which yields
\begin{align}
    c^\dagger_{\bk + \bG} c_{\bk + \bq + \bG} \lambda_\bq(\bk + \bG) &= c^\dagger_{\bk} c_{\bk + \bq } \lambda_\bq(\bk) e^{i \frac{B}{2} \bG \wedge [\bk] + i \frac{B}{2} \{\bk \} \wedge \bar \bG} e^{-i \frac{B}{2} \bG \wedge [\bk + \bq] - i \frac{B}{2} \{\bk + \bq\} \wedge \bar \bG} e^{-i \frac{B}{2} \bq \wedge \bG } \nonumber \\
    &= c^\dagger_{\bk} c_{\bk + \bq } \lambda_\bq(\bk) e^{-i \frac{B}{2} \{\bk + \bq\} \wedge (\bG + \bar \bG)} = c^\dagger_{\bk} c_{\bk + \bq } \lambda_\bq(\bk) e^{- 2\pi i C \{\bk + \bq\}_2 G_1 } = c^\dagger_{\bk} c_{\bk + \bq } \lambda_\bq(\bk) 
\end{align}
On going from the first to the second line, we have assumed $\bk$ lies in the first BZ such that $\{\bk\} = 0$. In the second inequality we wrote the  $\bG = G_1 \bb_1 + G_2 \bb_2$ and $\{\bk + \bq\} = \{\bk + \bq\}_1 \bb_1 + \{\bk + \bq\}_2 \bb_2$ and used $B \bb_1 \wedge \bb_2 = - B A_{\rm BZ} = -2\pi$. Finally, we used the fact that $\{\bk + \bq\}_{1,2}$ and $G_{1,2}$ are integers. 

Let us define $\tilde \rho_\bq = \rho_\bq e^{\frac{|B|}{4} \bq^2}$  which satisfy the modified GMP algebra
\begin{equation}
    [\tilde \rho_\bq, \tilde \rho_{\bq'}] = 2 i \sin \left(\frac{B}{2} \bq \wedge \bq' \right) \tilde \rho_{\bq + \bq'}
\end{equation}
With this definition, we find that
\begin{equation}
    \tilde \rho_\bq = \sum_{\sigma,\bk \in {\rm BZ}} c^\dagger_{\sigma,\bk} c_{\sigma,[\bk + \bq]} \tilde \lambda_\bq(\bk) = \sum_{\sigma,\bk \in {\rm BZ}} \tilde c^\dagger_{\sigma,\bk} \tilde c_{\sigma,\bk + \bq} \tilde \lambda_\bq(\bk), \qquad \tilde \lambda_\bq(\bk) = e^{-i \frac{B}{2} \bq \wedge \bk} e^{i (\phi_{\bk + \bq} - \phi_\bk)}
    %\quad \tilde \lambda_\bq(\bk) = e^{-i \frac{B}{2} \bq \wedge \bk - i \frac{B}{2} (\{\bk + \bq\} \wedge [\bk + \bq] - \{\bk\} \wedge [\bk]) - i \frac{B}{4} (\{\bk + \bq\} \wedge \{\bar \bk + \bar \bq\} - \{\bk\} \wedge \{\bar \bk\})}
\end{equation}
In the following, we will assume that the gauge is periodic such that $ c_\bk = c_{[\bk]}$.

The spin waves in the model can be understood by defining the spin raising operator
\begin{equation}
    \tilde S_\bq^+ = \frac{1}{\sqrt{N}} \sum_\bk c_{\uparrow, \bk}^\dagger c_{\downarrow, \bk + \bq} \tilde \lambda_\bq(\bk) = \widehat{\tilde \lambda_\bq \otimes \sigma^+}, \qquad \sigma^+ = \frac{1}{2} (\sigma_x + i \sigma_y)
\end{equation}
It is easy to verify that the commutation relation
\begin{equation}
    [\delta \tilde \rho_\bq, \tilde S^+_{\bq'}] = 2 i \sin \left(\frac{B}{2} \bq \wedge \bq'\right) \tilde S^+_{\bq + \bq'}
\end{equation}
which implies that $|\bq \rangle = \tilde S_\bq^+|\downarrow \rangle$ is an exact eigenstate of $\H$ (note that for the LLL, the single particle term $\epsilon_0(\bk)$ vanishes) since
\begin{equation}
    \H |\bq \rangle = \xi_{\bq} |\bq \rangle, \qquad \xi_{\bq'} =  \frac{2}{A} \sum_{\bq'}  \tilde V_{\bq'} \sin^2 \left(\frac{B}{2} \bq \wedge \bq'\right) \simeq 2 \rho \bq^2 + O(\bq^4), \qquad \rho =  \frac{B^2}{8A} \sum_{\bq'}  \tilde V_{\bq'} {\bq'}^2 = \frac{B^2}{16\pi} \int dq q^3 V_q e^{-\frac{B}{2} q^2}
\end{equation}
In addition $|\bq \rangle$ satisfy
\begin{equation}
    \langle \bq'| \bq \rangle = \frac{1}{N} \sum_{\bk,\bG} \delta_{\bq,\bq' + \bG} \tilde \lambda^*_{\bq'}(\bk) \tilde \lambda_{\bq}(\bk) = \frac{1}{N} \sum_{\bG} \delta_{\bq,\bq' + \bG} e^{- i \frac{B}{2} \bG \wedge \bq - i \frac{B}{2} \bG \wedge \bar \bG} \sum_\bk e^{-i B \bG \wedge \bk} = \delta_{\bq, \bq'}
\end{equation}
where $\bq$ ranges over the linearly indepdent generators of the GMP algebra given by $\bq = q_1 \bb_1 + q_2 \bb_2$, $0 \leq q_1 < N_1$, $0 \leq q_2 < N_2$. Note that this is different from the case of a general Chern band where $\bq$ is defined to live only within the first BZ with an additional index $n$ that goes from $1$ to $N$. Here instead $\bq$ itself ranges over $N^2$ points in the extended BZ.

We define the states
\begin{equation}
    |\bk_0; \bq \rangle = \tilde \lambda^*_\bq(\bk_0) c_{\uparrow,\bk_0 + \bq}^\dagger \tilde S^+_{\bq} |\downarrow \rangle
\end{equation}
which satisfy
\begin{align}
    \langle \bk_0'; \bq' |\bk_0; \bq \rangle &= \frac{1}{N} \tilde \lambda_{\bq'}(\bk'_0) \tilde \lambda^*_\bq(\bk_0) \sum_{\bk,\bk'} \tilde \lambda^*_{\bq'}(\bk') \lambda_\bq(\bk) \langle c_{\downarrow, \bk' + \bq'}^\dagger c_{\uparrow,\bk'} c_{\uparrow,\bk'_0 + \bq'} c^\dagger_{\uparrow,\bk_0 + \bq} c^\dagger_{\uparrow,\bk} c_{\downarrow, \bk + \bq} \rangle \\
    &= \frac{1}{N} \tilde \lambda_{\bq'}(\bk'_0) \tilde \lambda^*_\bq(\bk_0) \sum_{\bk,\bk'} \tilde \lambda^*_{\bq'}(\bk') \lambda_\bq(\bk) [\delta_{\bk' + \bq', \bk + \bq} \delta_{\bk, \bk'} \delta_{\bk'_0 + \bq', \bk_0 + \bq} - \delta_{\bk' + \bq', \bk + \bq} \delta_{\bk, \bk'_0 + \bq'} \delta_{\bk', \bk_0 + \bq} ] \\
    &= \delta_{\bk_0, \bk'_0} [\delta_{\bq, \bq'} - \frac{1}{N} \tilde \lambda_{\bq'}(\bk_0) \tilde \lambda^*_{\bq}(\bk_0) \tilde \lambda^*_{\bq'}(\bk_0 + \bq) \tilde \lambda_\bq(\bk_0 + \bq')] 
\end{align}
where we assumed that $\bk_0$ and $\bk'_0$ lie in the first BZ. The action of $\H$ on $|\bk, \bq \rangle$ is given by
\begin{equation}
    \H |\bk_0; \bq \rangle = \xi_{\bq}  |\bk_0; \bq \rangle + \frac{i}{A} \sum_{\bq'} \tilde V_{\bq'} \sin \left(\frac{B}{2} \bq' \wedge \bq\right) (e^{i\frac{B}{2} \bq' \wedge \bq} |\bk_0 \bq + \bq' \rangle - e^{-i\frac{B}{2} \bq' \wedge \bq} |\bk_0; \bq - \bq' \rangle)
\end{equation}
As we can see, the Hamiltonian is independent on $\bk_0$ which implies that all bands, single particle and polaron, are perfectly flat. This is a consequence of contineous magnetic translation.

The bound state wavefunction can be written as
\begin{equation}
    |\Psi \rangle = \sum_\bq r_\bq |\bk_0; \bq \rangle
\end{equation} 
We reiterate here that $\bq$ is not restricted to the first BZ. The magnitude and phase of  corresponding to the bound state are shown in Fig.~\ref{fig:LLL} (this corresponds to unfolding the wavefunction in Fig.~\ref{fig:Wavefunctions}). We can see that the function decays exponentially away from $\bq$ with its phase winding by $2\pi$ around $\bq = 0$. This leads us to propose the variational ansatz
\begin{equation}
    a_\bq(\eta) = e^{-\frac{\xi}{2} |\bq| + i \phi_\bq}, \quad \phi_\bq = \arg(q_x + i q_y) \qquad |\psi_\eta \rangle = \sum_\bq a_\bq(\eta) |\bq,\bq \rangle
    \label{aq}
\end{equation}
The overlap of the ansatz wavefunction with the exact ground state as a function of $\xi$ (measured in units of $l_B$) is shown in Fig.~\ref{fig:LLL} and we can see that for the optimal value of $\xi$ around $\xi \approx 2.6$, we get an overlap exceeding 99\%. We can also see in Fig.~\ref{fig:LLL}, the minimum variational energy compared to the exact ground state energy and we see the two match very closely.

\begin{figure}
    \centering
    \includegraphics[width = 0.55 \textwidth]{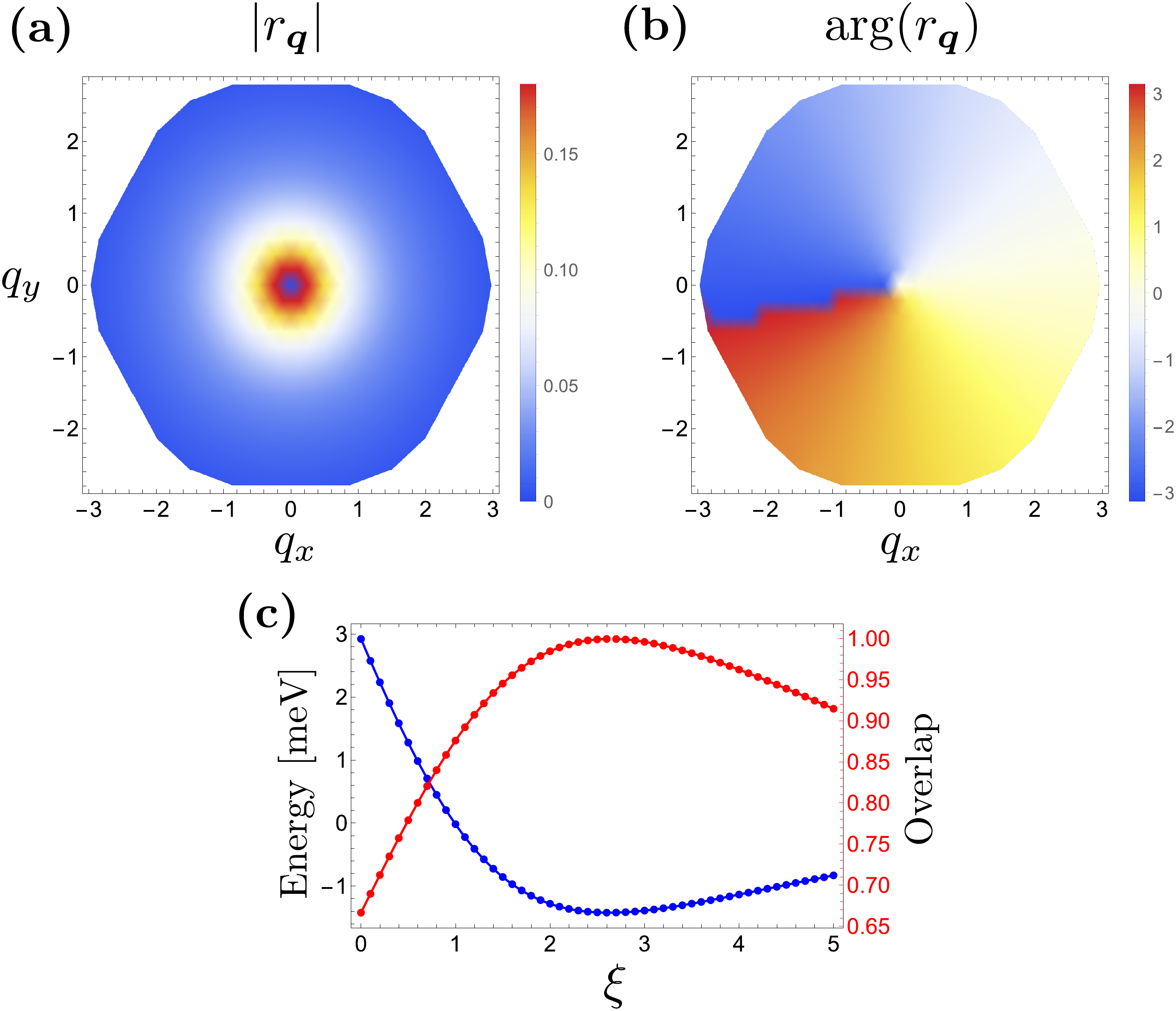}
    \caption{{\bf (a)} Mangitude of the bound state wavefunction, {\bf (b)} Phase of the bound state wavefunction, and {\bf (c)} Variational energy and overlap with the exact wavefunction as a function of the variational parameter $\xi$ measured in units of $l_B$.}
    \label{fig:LLL}
\end{figure}

\section{Dispersion}
Here, we discuss some details about the dispersion we used in the main text. At integer fillings and if we ignore the inter-Chern part of the dispersion (the so-called flat band limit), the single particle dispersion is given exactly \cite{KangVafekPRL,TBGV} by diagonalizing the Hartree-Fock Hamiltonian \cite{ShangNematic, KIVCpaper} which takes the form
\begin{equation}
    H_{\rm HF}[Q](\bk) = \frac{1}{2A} \sum_\bG V_\bG \Lambda_\bG(\bk) \sum_{\bk'}\tr \Lambda_{-\bG}(\bk') Q_\bk - \frac{1}{2A} \sum_\bq V_\bq \Lambda_\bq(\bk) Q_{\bk + \bq} \Lambda_\bq(\bk)^\dagger
    \label{HHF}
\end{equation}
Here, $Q_\bk$ is a matrix with eigenvalues $\pm 1$ describing  a Slater determinant state such that $\pm 1$ correspond to full/empty electronic states. $\Lambda_\bq(\bk)$ is a matrix for form factors with spin $(s)$, sublattice $(\sigma)$ and valley $(\tau)$ indices which can be transformed into a Chern $(\gamma)$, spin $(s)$ and pseudspin $(\eta)$ basis (see Refs.~\cite{KIVCpaper, SkPaper, KhalafSoftMode}). Since our analysis focuses on a single Chern sector, we are going to neglect the Chern off-diagonal terms in the form factor $\Lambda$ which was shown in Ref.~\cite{KIVCpaper} to be relatively small (they vanish identically in the chiral limit). In this limit, $\Lambda_\bq(\bk)$ takes the simple form $\Lambda_\bq(\bk) = s_0 \otimes \eta_0 \otimes \diag(\lambda_\bq(\bk), \lambda_\bq^*(\bk))_{\gamma}$ where $\lambda_\bq(\bk)$ are the form factors for a single Chern band. At integer filling $\nu$ and ignoring inter-Chern dispersion, the family of the ground state is described $Q_\bk$ can be chosen to be $\bk$-independent and to satisfy $\tr Q = 2\nu$ and $[Q, \Lambda_\bq(\bk)] = 0$ which is equivalent to the condition that $Q$ is Chern-diagonal, i.e. $[Q, \gamma_z] = 0$ \cite{KIVCpaper}. Under these conditions, the Hamiltonian simplifies to
\begin{equation}
    H_{\rm HF}[Q](\bk) = \epsilon_{\nu,\pm}(\bk) = \pm \epsilon_F(\bk) + \nu \epsilon_H(\bk), \quad  \epsilon_H(\bk) = \frac{1}{A} \sum_\bG V_\bG \lambda_\bG(\bk) \sum_{\bk'} \lambda_{-\bG}(\bk'), \quad \epsilon_F(\bk) = \frac{1}{2A} \sum_\bq V_\bq |\lambda_\bq(\bk)|^2
\end{equation}
where the positive (negative) sign is for the electron (hole) bands. We note that $\epsilon_{-\nu,\pm} = -\epsilon_{\nu,\mp}$ due to particle-hole symmetry so that electron (hole) bands on the $\nu > 0$ side map to hole (electron) bands on the $\nu < 0$ side. That is, doping away from charge neurality is the same whether for positive and negative $\nu$ and similarly for doping towards neutrality. The Hartree and Fock potentials are plotted in Fig.~\ref{fig:HF}a and we can see that both are characterized by a dip at $\Gamma$. Thus, for doping away from neutrality, the two are going to add while on doping towards neutrality they subtract. In the main text, we used $\nu$ as an interpolation parameter that also takes non-integer values as a proxy for tuning the bandwidth. %This can be justified by noting that the Hartree dispersion does indeed have the approximate form $\nu \epsilon_H(\bk)$ even away from integers (see Fig.~\ref{fig:HF})b if we introduce doping by gradually filling the band minimum. 
We can see in Fig.~\ref{fig:HF}b the bandwidth as a function of $\nu$ and we see that there is a minimum in the range $\nu \in [-1.5,-1]$ depending on the chiral ratio $\kappa$. The value of $\nu = \nu_{\rm min}$ for which the bandwidth is minimum is shown in Fig.~\ref{fig:HF}c. We note that for $\nu > \nu_{\rm min}$, the band minimum is at $\Gamma$ whereas for $\nu < \nu_{\rm min}$, the band maximum is at $\Gamma$. 

\begin{figure}
    \centering
    \includegraphics[width = 0.75 \textwidth]{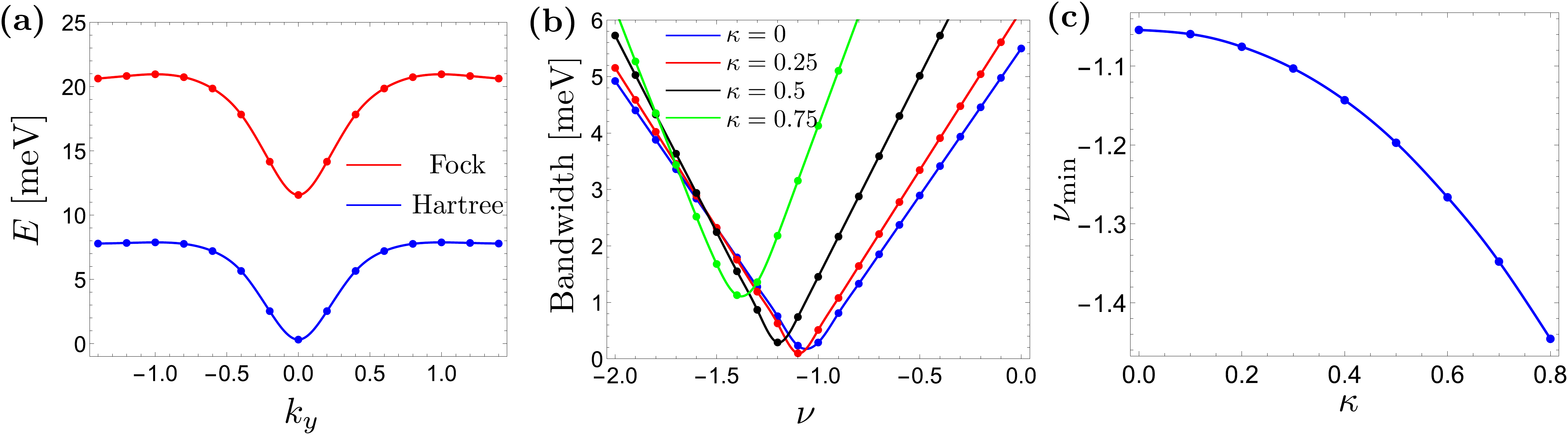}
    \caption{{\bf Details of the approximate Hartree-Fock dispersion}: {\bf (a)} Hartree $\epsilon_H(\bk)$ and Fock $\epsilon_F(\bk)$ dispersion. {\bf (b)} Bandwidth for the dispersion $\epsilon_{\nu}(\bk)$ as a function of $\nu$ for different values of $\kappa$. {\bf (c)} Value of $\nu$ for which the bandwidth is minimum as a function of $\kappa$.}
    \label{fig:HF}
\end{figure}

\end{document}